# Provide High-QoS of the High-Speed Railway Mobile Communications in Cyber-Physical Systems

Yuzhe Zhou
State Key Lab of Rail Traffic Control and Safety
Beijing Jiaotong University
Beijing, China
e-mail:12120198@bjtu.edu.cn

Bo Ai *(Corresponding Author)*
State Key Lab of Rail Traffic Control and Safety
Beijing Jiaotong University
Beijing, China
e-mail: boai@bjtu.edu.com

*Abstract*—Technical advances in networks, embedded computing, and wireless communications are leading to the next generation of complex intelligent systems called Cyber-Physical Systems (CPS). CPS promises to transform the way we interact with the physical world. Efficient and reliable operation of the CTCS-3 (Chinese Train Control System Level 3) is great protection of the national economy and public safety. The CTCS-3 is based on GSM-R (GSM for Railway) to achieve a continuous and two-way transmission of information between ground and the train. To ensure the growing needs of safety, fastness and service diversity of China's railway, the pursuit of high-QoS has been the key of the relative study.This paper examines the main characteristics of GSM-R and the requirements of all-layers' QoS indicators for GSM-R. Several main technologies of improving QoS indicators of all-layers are summarized. As a solution, a comprehensive scheme is proposed to improve the delay and packet loss indicators. An example is also presented that illustrates the real-time features of the proposed solution. Based on the CPS characteristics are highly correlated with the QoS indicators, conclusions are made that GSM-R can provide a reliable and real-time way for message.

I. INTRODUCTION

Cyber-Physical System is based on the technologies of networks and embedded systems and develops into the next generation of complex intelligent systems. CPS stresses the integration of computing, communications, controlling and physical systems [1]. It has self-adaptability to the change of environmental uncertainty, dynamic reconfiguration function and network-based integration control of large-scale systems.

Efficient and reliable operation of the CTCS-3 (Chinese Train Control System Level 3) is great protection of the national economy and public safety. The CTCS-3 uses the technology of digital mobile telecommunication based on the GSM-R (GSM for Railway) to achieve a continuous, two-way transmission of information between ground and the train. As the developing direction of China's private railway communication system, GSM-R needs to provide a fast and integrated transmission channel for voice and data communications of railway transportation, that is to say high-availability and high-reliability are required in GSM-R network. To ensure the growing needs of safety, fastness and service diversity of China's railway, a set of quality of service (QoS) system has been established, and still the pursuit of high-QoS has been the key of the relative study.

In this paper, we will apply the concepts of CPS to study high-QoS of the private mobile communication system for high-speed railway. The requirements of all-layers' QoS indicators for GSM-R are examined in Section 2. A set of main technologies of improving QoS indicators of all-layers is summarized in Section 3. Higher-layer switching technology is examined in Section 4 particularly. A comprehensive scheme of higher-layer switching technology is proposed to improve the delay and packet loss indicators of GSM-R in Section 5. An illustrative example is presented with promising results. Finally, concluding remarks are given in Section 6.

II. CPS CHARACTERISTICS AND GSM-R QoS REQUIREMENTS

In a future CPS, a large number of embedded, possibly mobile computing devices will be interconnected through wireless communication, constituting various autonomous subsystems that provide certain services for end users. Global information sharing is achieved by connecting to the Internet. For instance, in a cyber-physical system, there may be diverse cyber-physical subsystems like personal health care, smart home, intelligent transportation and public security. CPS will become pervasive in virtually all fields of science and engineering, such as industry, health care, building, security, environmental science and biology as well as our everyday life.

Compared to various existing smart technology (embedded systems, Internet of Things, etc.), CPS has the following main characteristics [2] which are closely related to QoS provisioning:

*A. Heterogeneity.*

Possibly designed using different technologies and with different goals, devices are different from each other in many aspects such as capabilities, functionality, and number. In a large-scale CPS, the hardware, software and networking technologies used in the system may differ from one subsystem to another. A variety of services are also mixed.

## B. Mixed traffic.

Diverse applications may need to share the same communication methods or different communication methods that induce both periodic and aperiodic data. Furthermore, devices for different kinds of physical variables, e.g. location and speed, generate traffic flows with different characteristics (e.g. message size and sampling rate). Different applications will have different characters.

## C. Dynamic network.

Node mobility is an intrinsic nature of many applications such as intelligent transportation. During runtime, new nodes may be added or removed; some nodes may even die due to exhausted battery energy. All of these factors cause the network topologies of CPS to change dynamically and sometimes quickly.

## D. Resource constraints.

Network nodes are usually low-cost, low-power, small devices that offer only limited data processing capability, bandwidth, transmission rate, battery energy, and memory. In particular, energy conservation is critically important for extending the lifetime of the system. Dynamic and rapid management of resources is required.

CPS is application-oriented by nature. Therefore, networks have to provide QoS support so as to satisfy the service requirements of target applications. From an end user's perspective, real-world applications have their specific requirements on the QoS of the underlying network infrastructure. For instance, in the railway control system, sensors need to report the occurrence of a coming accident to actuators in a timely and reliable fashion; then, the actuators will react by a certain deadline so that the situation will not become uncontrollable.

CTCS-3 as a subsystem of CPS is a train operation control system based on wireless communications—GSM-R. Starting from the GSM-R system to determine the QoS indicators is significant to provide highly reliable and efficient wireless network for China's railway system, which is for CPS.

Different applications may have different QoS requirements. For instance, for a safety-critical control system, large delay in transmitting data from control center to the train and packet loss occurring during the course of transmission may not be allowed, while they may be acceptable for a mailing system that used in a train.

QoS can be regarded as the capability to provide assurance that service requirements of applications can be satisfied. Depending on the application, QoS in GSM-R can be characterized by reliability, timeliness, robustness, availability, and security, among others. Some QoS parameters may be used to measure the degree of satisfaction of these services, such as throughput, delay, jitter, and packet loss rate.

The requirements of the QoS indicators of GSM-R for train control message transmission (TCMT) service [3] are shown in Table 1 including end-to-end transmission delay ($T_d$), transmission interference time ($T_{TI}$), transmission error-free time ($T_{REC}$), etc.

TABLE I.   GSM-R NETWORK TRAIN CONTROL DATA SERVICE QOS INDICATORS

| QoS Item | Indicator Value |
|---|---|
| Connection Establishment Delay | $< 8.5s(95\%), \leq 10s(100\%)$ |
| Connection Establishment Failed Possibility | $< 10^{-2}$ |
| Connection Failure Rate | $\leq 10^{-2}$ |
| $T_d$ | $\leq 0.5s(95\%)$ |
| $T_{TI}$ | $< 0.8s(95\%), < 1s(99\%)$ |
| $T_{REC}$ | $> 20s(95\%), > 7s(99\%)$ |
| Network Registration Delay | $\leq 30s(95\%), \leq 35s(99\%), \leq 40s(100\%)$ |

On improving the reliability, on the one hand, can be achieved by enhancing the real-time ability of the system; on the other hand, can make use of some of the existing network invulnerability and cascade accident prevention technology to study the prevention of sudden abnormal events, and can achieve real-time forecasting and recovery techniques. We must also ensure that the energy of the system to maintain long-lasting, such as through coordination between the various components and scheduling to achieve energy saving, uninterruptible power supply, or R&D of new energy storage for longer life devices.

## III. MAIN TECHNOLOGIES OF IMPROVING QOS INDICATORS

GSM-R network is the communication platform of the CTCS-3, the GSM-R QoS indicators must meet the relevant requirements of the CTCS-3. Similarly, the CTCS-3 also has its own QoS indicators to meet the operational needs of China's railway transportation [4], and railway transportation is ultimately serves for customer. QoS indicators in each layer are not existing in isolation, but influencing each other. So the train operating system should from user experience point of view and according to their business need to define and arrange CTCS-3 QoS indicators and GSM-R QoS indicators from top to bottom (i.e. top-level design). Only determined by these indicators, can we optimize the various parameters in the data transfer protocol. Thus the entire network can be better adapted to the QoS characteristics, as well as to meet the requirements of CPS characteristics.

High Speed Rail private mobile communication system has its unique train control QoS specification. In this section, we will analyze optimization techniques for delay and packet loss QoS indicators which are associated with real-time and reliability of the CPS system by the protocol stack. The protocol stack architecture of OSI [5] is shown in Fig. 1.

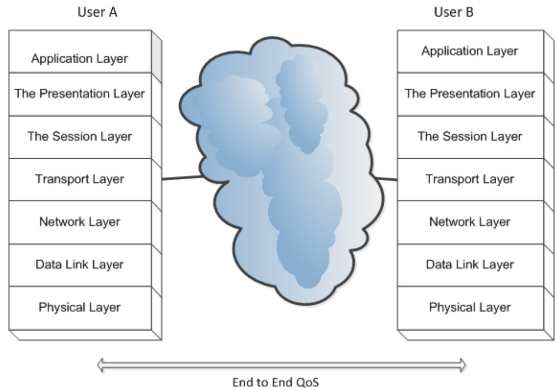

Figure 1. The protocol stack architecture

## A. Physical Layer and Data Link Layer

### 1) Interference analysis

The interference of the mobile communication system is an important factor affecting the quality and transmission reliability of the wireless network. Interference will cause the bit error to increase, voice quality decrease. In order to ensure the mobile communication network to operate effectively, so that all users are able to communicate without interference, the assessment and coordination for interference must be done. With different inhibition measures to deal with all kinds of interference, paper [6] introduced a number of ways: frequency configuration, the use of antennas, equalization techniques, diversity techniques and parameters configuration etc.

### 2) Handoff

The reliability and efficiency of the handoff is an important indicator that reflects the GSM-R system performance, and plays an important role in the protection of high reliability and security for railway mobile communication. The goal of handoff is to maximize the success rate of switching, reduce switching delay and improve the switching efficiency. Paper [7] analyzes the call drop of handoff in-depth and proposed a GSM-R network handoff switching point, interference sources and transmission loop optimization scheme to improve the stability of the handoff and ensure train's steady, high-speed and safe operation.

### 3) Real-time protocol

Taking each layer into account, establish various real-time mechanisms. Paper [8] shows the problem finding and signaling transmission mechanism of the mobile networks in the physical layer, link layer and TCP/IP layer. Paper [9] designed a real-time protocol based on the original protocol which does not modify the existing protocol stack but add a new real-time protocol layer, the scheme is as follows:

*a)* RTMAC protocol layer added and the MAC protocol layer removed, traffic smooth processing technology used to limit the non-real-time streaming as much as possible, while limit the real-time streaming as little as possible.

*b)* Based on the traffic smoothing technique, new real-time communication protocol stack added.

### 4) Seamless network integration

Fixed Mobile Convergence (FMC) technology includes a variety of business, terminal and network integration. Wireless technology like GSM, Bluetooth, and Wi-Fi are applied. Integrated device is a kind of intelligent device which is able to connect multiple networks to provide fixed and mobile network connectivity. From the network access point of view, to achieve the FMC in two ways: one is by way of access to the fixed network; the other is through access to mobile networks. From the user's perspective, the goal of FMC is to provide users with complete, seamless, consistent communications service experience, allowing users to get rid of the limitations of space and network diversity.

## B. Network Layer

### 1) Anycast

Common routing algorithm is usually based on the distance of the routing path, the packet will be sent to an address of the nearest mirror server, and however, the nearest mirror server is not always the best mirror server. This kind of shortest path routing is likely to cause network congestion, make the network load unbalancing, decrease network performance. In order to enhance the availability of services and improve the distribution of network traffic, the anycast communication model is proposed. Anycast means to elect a best service node to provide services for user's QoS request. Paper [10] proposed a new heuristic algorithm to solve the multiple constraints QoS routing problem, considering the delay, bandwidth and cost constraints. The algorithm is effective to solve the QoS routing problem under a variety of restrictions, and achieved a better balance of the network load and server load.

### 2) Real-time network processing

At some point, a node can receive multiple packets need to deal with, or there are multiple packets need to be sent out, which packet should be processed or sent firstly? Currently, the best-effort network is based on FCFS manner, but this does not comply with the requirements of real-time systems like CPS. Paper [9] proposed a priority associated with the degree of importance or urgency of the information to each packet. It gives high-priority packet to be processed or sent at first. In order to reduce the queuing time before processing or sending the high-priority packets, and enhance the predictability of the high-priority packets' delay in the network transmission and thus improve the real-time transmission of the network.

### 3) Power control

Traditional routing protocols use a set of indicators to find and maintain the optimal routing path, the most commonly used indicator is the least number of hops. However, this indicator does not take the energy and life of the node and network into account. This may lead to problems like energy waste and excessive use, and ultimately make the life span a serious decline of the nodes and networks. Paper [11] introduces an energy-aware strategy which is to make the remaining energy of the path to become a major indicator to select the optimal routing path. Choosing the path of the maximum residual energy to transmit data, and reduce the node energy waste and overuse. Based on the wireless network topology, paper [12] proposed an energy consumption management method for nodes within the network, expecting to reduce energy consumption and protect the real-time and priority constraints at the same time.

*C. Transport Layer to Application Layer*

The mechanism of QoS work is to give difference treatment to different data streams. In order to ensure the transmission of critical data streams to get the best services, the mechanism should protect some data stream, and at the same time limit other data flow. Nowadays, growing range and needs of services ask for greater efficiency and higher QoS data exchange technology. It can be seen that under the current exchange system which is mainly in layer 2-3 is very difficult to achieve the QoS guarantee. There is no way to guarantee the high QoS requirements of data streams of special treatment. In order to solve the QoS issue, a higher layer switching technology [13] had been proposed based on the traditional layer 2-3 exchange technology. The higher layer switching technology can guarantee user satisfaction with specific services while improve the QoS parameters. Higher layer switching technology applied to system is a kind of intelligent switching technology for data stream control and processing.

Applying of the above technologies can achieve the optimization of the system QoS indicators, allowing the system to reflect the real-time, reliability, heterogeneity, integrity and other characteristics of the CPS.

## IV. HIGHER LAYER SWITCHING TECHNOLOGY

The transmission of data packets in the $4^{th}$ layer switch is not only based on MAC address ($2^{th}$ layer), source/destination IP address ($3^{th}$ layer), but also based on TCP/UDP port address ($4^{th}$ layer) [14]. The port addresses represent different service protocols, that is to say the 4th layer switching technology is not only a physical exchange, but also includes service exchange. The $4^{th}$ layer switching technology supports intelligent features like network traffic and QoS control which lower layer switching technologies cannot possess. The $4^{th}$ layer switching technology with service intelligence can make the decision where to forward the session transport stream. It can ensure that data flow smoothly between the client and server, achieve a nice balance to the load of all networked devices, and live up to end-to-end QoS requirements. Selection strategy in the $4^{th}$ layer switch should be formulated according to different needs. To ensure the QoS of high priority service first, and lead this service to the most available server.

The $7^{th}$ layer switching technology makes exchanges into the range of process and content [15]. As related with applications, the exchange should have intelligence. The switch has the ability to distinguish various high layer applications and identify contents. In this case the switches not only send data according to the packet's IP address or port address, but also take the package information into account by open the packet and go inside the packet. This technology can make load balancing, content recognition and judgment, and make more smart decisions of the streams based on the application types. The $7^{th}$ layer switch allows users to make decisions freely to all kinds of transport streams and their destination based on the information obtained. Thus it optimizes network access and provides a better service for end users. This will truly realize the QoS requirements of CPS.

## V. A COMPREHENSIVE SCHEME

This section will propose a comprehensive scheme as a solution for realizing some of the functions identified in Section 4.

Multilayer switching technology's stream classification function can analyze the port address, IP address, MAC address, ToS (Type of Service) of the network data stream classification synthetically. On this basis, functions such as flow control, setting the priority, exchange redirection etc. are realized to use network bandwidth reasonably and support high-QoS requirements.

The vision of the model structure is shown in Fig. 2: data packets will be sent to the stream classification module for processing according to the MAC address, receiving ports or routing information in them. Then stream classification module will query the stream classification table based on the information in the packets. The packet stream classification rules operate such as changing the destination port number when the entry matches. Firstly, the data stream processes layer-2 switching, layer-3 routing is required if the packages have to be sent inter-network, data stream can be further processed with the protocol field and source/destination port number information. This scheme takes all needed information into account. It will surely provide high-QoS requirements.

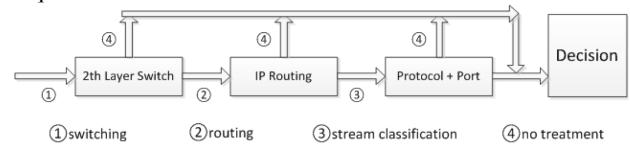

Figure 2. A comprehensive scheme's model structure

*A. An application example*

To illustrate the feasibility of the above scheme, consider a simple as shown in Fig. 3, where bottlenecks are PPP_E1 (2.048Mbps) links, others are 10Base_T (10Mbps) links, it has 2 servers, 3 routers, 4 switches, 8 clients with various applications like E-mail, http, voice, video conference etc. We make this model to simulate the high-layers of the GSM-R system simply.

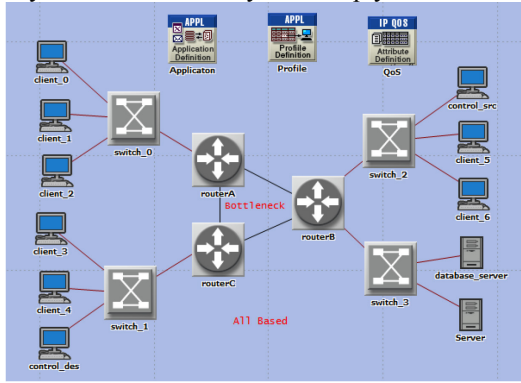

Figure 3. A simple network model

The configuration is as follows. Firstly, we set E-mail, http, voice and video conference these applications with ToS of BF, BF, EF and AF respectively. Then, we also configured profiles for each client according to the application configuration. Finally, we configured 4 QoS schemes which are ToS based, Protocol based, port based and all based (comprehensive scheme), together with a scheme without QoS awareness.

Fig. 4 depicts the packet loss for each scheme in video conference application. It is clear that the packet loss is dramatically in schemes but port based and all based schemes.

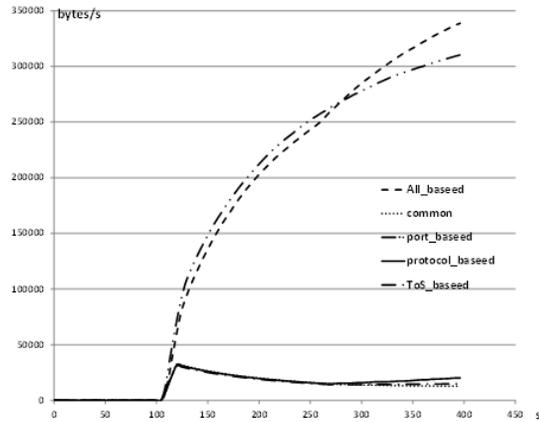

Figure 4. Packet loss in video conference application

Fig. 5 shows the delay for each scheme in different applications. It can be seen from Fig. 5-(a) shows that the delay for 5 schemes in voice application are around 3s (ToS based scheme), 1.5s (non-QoS and protocol based schemes), less than 1.5s (port based and comprehensive schemes) respectively. Fig. 5-(b) shows that the delay for 5 schemes in video conference application are around 4s (non-QoS and protocol based schemes), 3s (ToS based scheme), 0.5s (port based and comprehensive schemes) respectively.

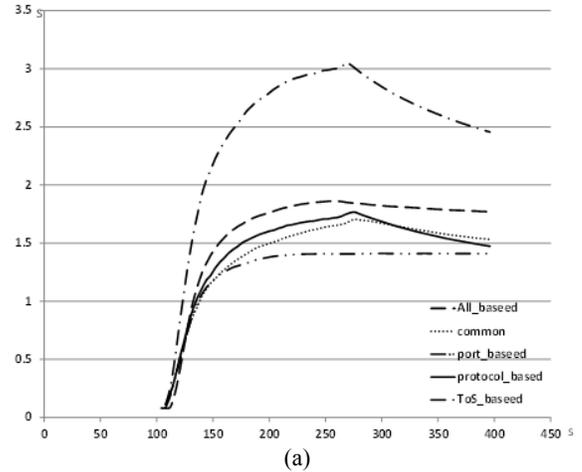

(a)

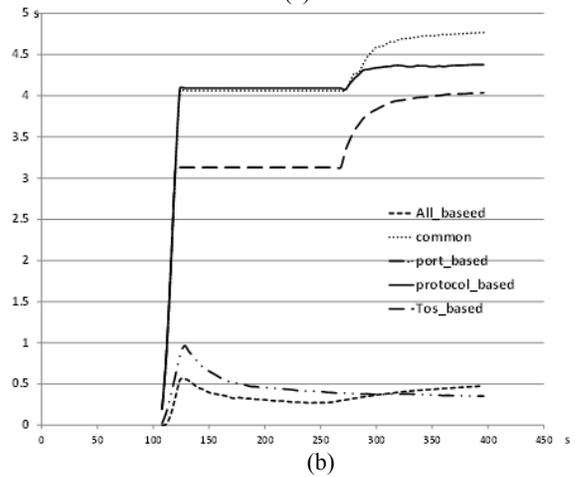

(b)

Figure 5. Delay of applications

Considering the length of the paragraph, we don't list all the results. Take all the results of the applications into account. Transference of large amounts of data, only port based and the comprehensive schemes can ensure better real-time and reliability. It can be seen, QoS scheme based on port number has good service applicability, real-time and reliability as well as the comprehensive scheme.

From these results, it can be summarized that the proposed higher layer switching technology compared to traditional switching technology can improve the real-time and reliability of high-priority applications without sacrificing real-time and reliability of low-priority applications in GSM-R, enhance the overall real-time and reliability indicators of the system, provide a solution for high-QoS requirements of the system. Finally, GSM-R can provide a reliable and real-time communication method for the CTCS-3 which is characterized by CPS. However, much work is left to be done in order to make it practically useful.

VI. CONCLUDING REMARKS

CPS is opening up unprecedented opportunities for research and development in numerous disciplines. Meanwhile, researchers currently face a large number of challenges that need to be overcome before the envisioned CPS become reality. This paper has examined the main characteristics of GSM-R and the requirements of its QoS indicators. Several main technologies of improving QoS indicators of all-layers are summarized. A comprehensive scheme of higher-layer switching technology is proposed as a solution to improve the delay indicator of GSM-R so the CTCS-3 can be more robust as a CPS subsystem. A simulation also verified that the real-time features of the proposed solution.

Extensive and deep research is expected in QoS technology for GSM-R and CPS.

### ACKNOWLEDGMENT

The authors would like to express their great thanks to the support from the 863 Plan of China under Grant 2011AA010104, the Fundamental Research Funds for the Central Universities under Grant 2010JBZ008, Program for New Century Excellent Talents in University under Grant NCET-09-0206 and the Key Project of State Key Lab under Grant RCS2011ZZ008.